# LONG-LIVED FILAMENTS IN FUSION PLASMAS: REVIEW OF OBSERVATIONS AND STATUS OF HYPOTHESIS OF MICRODUST-ASSEMBLED SKELETONS

A.B. Kukushkin and V.A. Rantsev-Kartinov

INF RRC "Kurchatov Institute",
Moscow, 123182, Russia

## ABSTRACT

A brief review is given of the progress in studying the long-lived filaments (LLFs) in fusion plasmas (tokamaks, Z-pinch, plasma focus). The paper reports on (i) resolving the fine structure of LLFs reported in [1], and (ii) verification of the hypothesis [1,2] which suggested the LLFs to possess a microsolid skeleton which might be assembled during electric breakdown, well before main plasma's birth, from wildly produced carbon nanotubes (or similar nanostructures of other chemical elements). The recent proof-of-concept studies showed the presence of tubular and cartwheel-like structures (i) in various types of dust deposits in tokamak T-10, in the range ~10 nm to ~10 μm, (ii) in the high-resolution visible light images of plasma, in ~100 μm to ~10 cm range, at initial stage of discharge (e.g., before appearance of discharge electric current) in tokamak, plasma focus and vacuum spark. The topological similarity of the above structures (especially, of the cartwheel as a structure of essentially non-hydrodynamic nature), and the observed trend of assembling bigger tubules from smaller ones (i.e. the self-similarity), allow to draw a bridge between the microdust skeletons in the dust deposits and the centimeter scale LLFs in fusion plasmas.

## 1. INTRODUCTION

The present paper gives a brief review of the progress in studying the long-lived filaments (LLFs) in fusion plasmas (tokamaks, Z-pinch, plasma focus) that have been achieved since last Symposium. Here we address the following two main issues.

First, we outline the major phenomenology of LLFs in fusion plasmas (Sec. 2). First of all, we present the data in which the fine structure of LLFs is resolved to a degree which proves definitely the tubularity of LLFs claimed in [1] on the basis of the less detailed data. Further, we show a new typical block which is essentially of a non-hydrodynamic nature, namely the cartwheel-like structure, i.e. a ring-shaped structure, with straight radial bonds connecting the ring with either an axle-tree (i.e. a straight filament) or a massive central point. The cartwheels are located in the edge cross-section of a tubule or exist as a separate block.

Second, we report on the proof-of-concept studies aimed at verification of the hypothesis [1,2] which suggested the LLFs to possess a microsolid skeleton which might be assembled during electric breakdown, well before major plasma's birth, from wildly produced carbon nanotubes (or similar nanostructures of other chemical elements). Here, we start with an outline of major hypotheses [1,2] suggested for interpreting the phenomenology [1] of LLFs (Sec. 3) and proceed (Sec. 4) with presenting the results of recent experiments (electron micrography of various types of dust deposits in tokamak T-10, laser shadowgraphy of initial stage of a vacuum spark discharge) and of analyzing the available databases from former experiments (high-resolution visible light imaging of plasma at electric breakdown stage of discharge in tokamak T-6 and plasma focus). The presence of tubular and cartwheel-like structures in the ranges ~10 nm to ~10 μm (in dust deposits) and ~100 μm to ~10 cm (in the plasma images), the topological similarity of these structures, and the observed trend of assembling bigger tubules from smaller ones (i.e. the self-similarity), allow to draw a bridge between the microdust skeletons in the dust deposits and the centimeter scale LLFs in fusion plasmas.

## 2. MAJOR PHENOMENOLOGY

Major phenomenology of LLFs in fusion plasmas may be reduced to the following two points.

First, an anomalously high survivability of the filaments, and their networks, was found in a broad range of laboratory plasmas that included

- gaseous Z-pinch [1,2,3(a,d),5(b)],
- tokamaks [3(a,b)],
- plasma foci [3(d)],
- laser-produced plasmas [1].

The anomalous survivability means that the filaments of the enhanced (or sometimes diminished) luminosity possess two significant properties: namely,

(i) regular geometric form (the elementary and most interesting block appears to be a straight tubular formation which often possesses a coaxial structure),

(ii) the lifetime is comparable with the entire duration of the electric discharge (this appears to exceed the existing theoretical predictions for such formation in respective experimental conditions by at least two, and more, orders of magnitude, especially for straight filaments directed perpendicular to main electric current; cf. Fig. 2 in [1] or Fig. 1 in [2(b)]).

The above properties made it worth to call such structures the *long-lived* filaments (LLFs) to distinguish them from the widely known phenomenon of a chaotic, short-lived filamentation.

Second, the *topological similarity* of such structures (including that of tubular structures) was found in a very broad range of

(i) length scales, from laboratory to cosmic space [1],

(ii) type of plasma confinement: namely, magneto-inertial (gaseous Z-pinches and plasma foci), magnetic (tokamaks), inertial (laser-produced plasmas).

Sometimes the self-similarity was found (i.e. the structure is build up by the topologically similar ones but of smaller, or much smaller, size).

In what follows we illustrate the most of above points successively for tokamaks, Z-pinch and plasma focus.

The analysis of all available databases was carried out with the help of the method of multilevel dynamical contrasting (MDC) [1,4,5(a)] (sometimes the large scale structuring may be seen without any processing, only a proper magnification of the image is needed). The reliability of the results is based on the very rich statistics, considerable similarity of the structures observed in various regimes and facilities, as well as on the insensitivity to specific way of imaging.

The typical examples for a number of tokamaks (mostly, small ones) are shown in Figures 1-7. The major parameters of these tokamaks (TM-2, T-4, T-6, T-10) are as follows: major radius, R (m) = 0.4, 0.9, 0.7, 1.5; minor radius, a (cm) = 8, 20, 20, 33; toroidal magnetic field, $B_T$ (T) = 2, 4.5, 0.9, 3; total plasma current, $I_p$ (kA) = 25, 200, 100, 300; electron temperature in the core, $T_e(0)$ (keV) = 0.6, 3, 0.4, 2; electron density in the core, $n_e(0)$ ($10^{13}$ cm$^{-3}$) = 2, 3, 2, 3.

The pictures 1-7 are taken in the visible light with the help of a strick camera (Figs. 1-4, 6-8) and a high-speed camera (Fig. 5). Everywhere the toroidal direction is the horizontal one. The effective time exposure for the strick camera is less than one microseconds (for the respective scheme of imaging see [6]). All the images correspond to plasma self-emission, except for Figure 5 where the light emitted by an injected pellet, managed to illuminate the LLFs in a broad spatial region.

The major features of the structuring are as follows:

(a) the length scale of the regular structuring varies in a broad range, from comparable with the minor radius of a tokamak to smallest resolvable lengths, i.e. less than millimeter scale (significantly, the presence of the large-scale structures proves the structuring to be present in a hot plasma interior, see Fig. 7(a));

(b) the typical tubule seems to be a cage assembled from (much) thinner, long rectilinear rigid-body structures which look like a solid thin-walled cylinders; often the cage takes the form of a few nested cages;

(c) the (almost rectilinear) tubules seem to form a network which starts at the farthest periphery (cf. Figs. 1,2) and is assembled from the tubules of various directions (e.g., toroidally directed tubules interconnected by those of radial and poloidal directions);

(d) a radial sectioning of the above network is resolved which looks like a distinct heterogeneity at some magnetic flux surface(s) (such a sectioning has been suggested [5(c)] to be related to the observed internal transport barriers in tokamaks).

Similar picture of LLFs appears to be found in plasma of Z-pinch experiments described in [4] and experiments on the plasma focus facility [7] (see also [8]). Here, our major attention was paid to identification of straight filaments directed nearly radially (see Figs. 8-11). The typical straight filament appears to be a cylindrical formation varying in length from few millimeters up to the radius (and even diameter) of the chamber. Such a filament has an axisymmetric tubular sheath, with a distinct boundary and, often, a distinct inner cylinder that makes the entire formation very similar to a coaxial cable (see, e.g., central window in Fig. 9, and Fig.11 ).

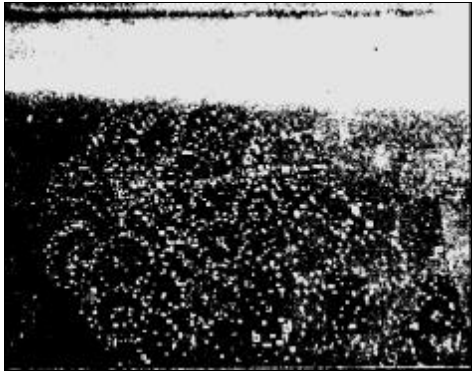

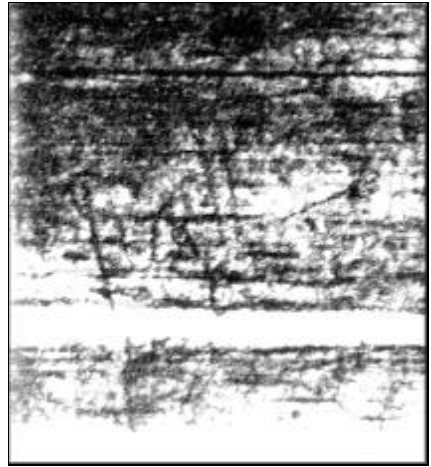

Figure 1. Circular structures in the limiter shadow in tokamak T-6 (minor radius a = 20 cm). Toroidal direction is horizontal one. The plasma column is seen as a white band on the top. Positive, image's height 10 cm. Diameter D of circles is ~1-1.5 cm, diameter d of the central spot inside circles is ~2-3 mm.

Figure 3. Long tubular filaments (D ~1-2.5 cm) in tokamak T-4 (a = 20 cm). Toroidal direction is horizontal one. Plasma core is on the image's top. Negative, figure's height 20 cm. (Thick horizontal white band in the lower part of the figure is a shadow of the reference wire located outside the chamber.)

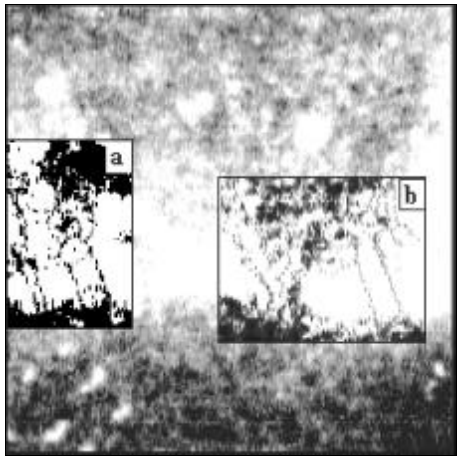

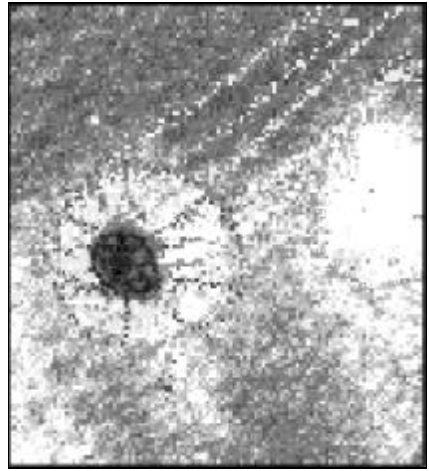

Figure 2. The lower half of the plasma column in tokamak T-6 (a = 20 cm). Positive, image's height 20 cm. Limiter shadow region is seen on the image's bottom. A tubule (D ~ 1 cm, with central spot ~ 2 mm) is seen in the window 'a' of the enhanced contrasting. The networking of similar tubular formations is seen in the window 'b'.

Figure 4. Upper half of plasma column in tokamak TM-2 (a = 8 cm). Positive, image's height 6.4 cm. Toroidal direction is horizontal one. Tubular structure (D ~ 2.5 cm) with a cartwheel in the edge cross section. Diameters of the inner dark circle ~1 cm, and that of a darker spot inside this circle ~3 mm.

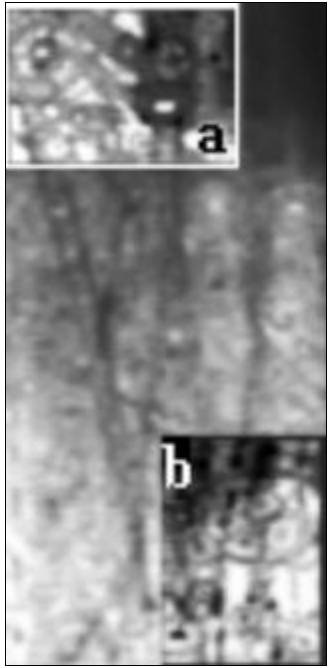

Figure 5. Tubular long filaments and cartwheel structures are seen in the visible light in the far periphery of tokamak T-10 (a = 33 cm) when illuminated by the carbon pellet emission (the pellet's track is outside the image). Negative, image's height is ~8.5 cm. Diameter of the long thick filament is ~ 3-4 mm. The windows "a" and "b" and the residual part of the image have different levels of contrasting to show the continuity of structuring and the fine structure of the cartwheels (see, e.g., a cartwheel in the left upper corner of window 'a'). The cartwheel in the window 'b' is located in the edge cross-section of a vertically aligned tubular structure.

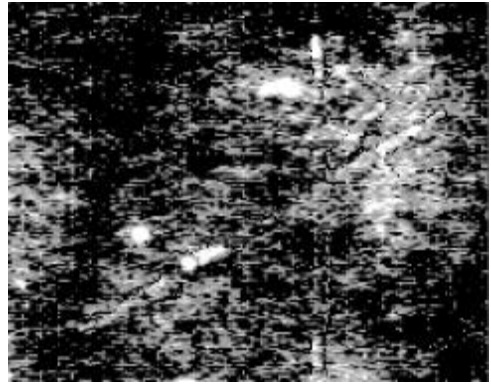

Figure 6. A rigid-body cross made of tubular filaments of diamater ~3-4 mm in tokamak TM-2 (a = 8 cm). Positive, image's height 6 cm.

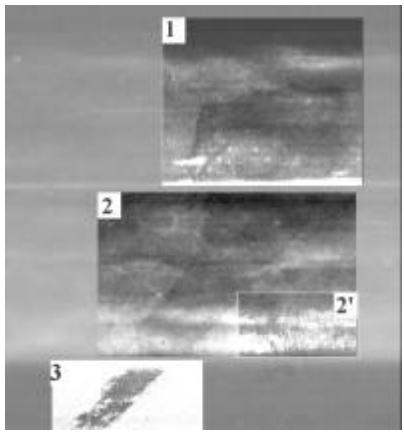

(a)

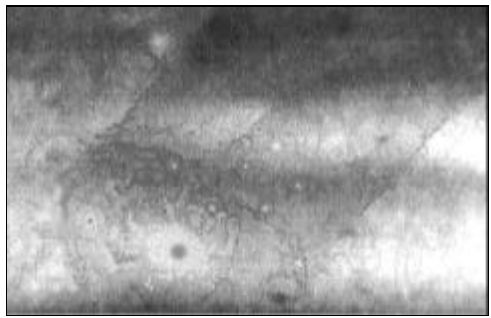

(b)

Figure 7. The extra-long tubular formation in tokamak TM-2 (a = 8 cm) is shown in Figure (a) (positive, image's height 20 cm). The tubule is stretched from the limiter-shadow region at the one side of the plasma column (i.e. outside the plasma column, see window 3) to similar region at its opposite side (dark horizontal band on the top of the window 1). The windows correspond to different levels (maps) of contrasting of the images, in order to show the continuity of the structuring. The tubular block seen on the bottom of the window 2 (namely, to the left from the window 2') is shown in Figure (b). Here, diameters of the tubule and central dark spot are ~2.5 cm and ~1.5 mm, respectively (image's height 2.8 cm).

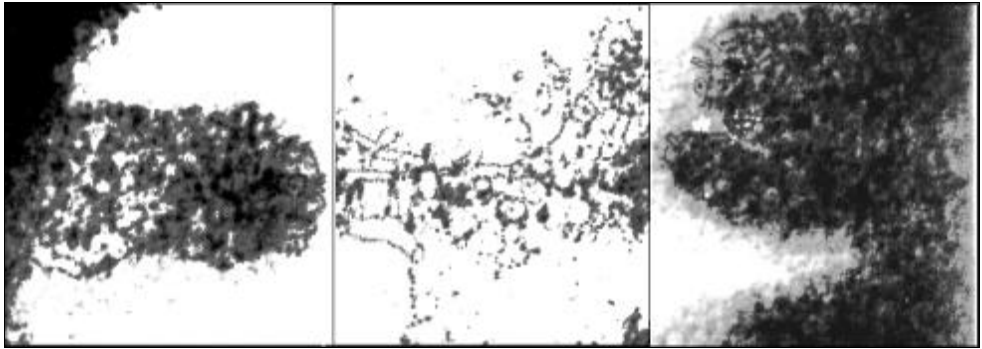

Figure 8. Visible light picture (positive) of a layer, 7.5 cm wide and 5 cm thick, located near the axis of a gaseous Z-pinch in the neck region at time t=+50 ns after major singularity of electric current (the layer is «extracted» by the optics collecting the light). The chamber is 60 cm long and 20 cm in diameter, major axis is directed vertically, maximal current ~ 360 kA, working gas deuterium, time exposure 10 ns (for other experimental conditions see [4,5(b)]). The original image is processed with the method of multilevel dynamical contrasting [4,5(a)] with different maps of contrasting in the central and peripheral windows to show the continuity of the structuring in the regions of substantially different luminosity. Dendritic tubular filaments (of a diminished luminosity, with respect to a stronger background) in the central section are of diameter d = 0.7 - 1.5 mm, while thick fractal formations («dark filaments») in the neck, in the left and right windows, are of d ~1.2 and 0.5 cm, respectively.

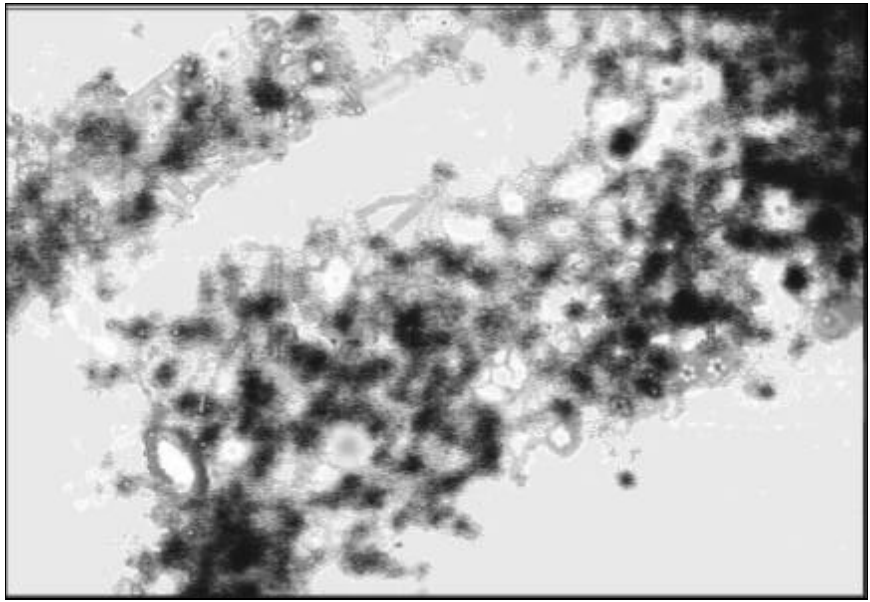

Figure 9. Typical radially-directed filamentary formation in a Z-pinch (major conditions are similar to those of Fig. 8). Here, the axis of the Z-pinch is located at the left edge of the image, time t =+300 ns, image's width 3.5 cm (positive), diameter of the ring at the left edge of a dark fractal filament is ~3 mm, and the thinnest resolvable tubules are few hundreds of microns in diameter.

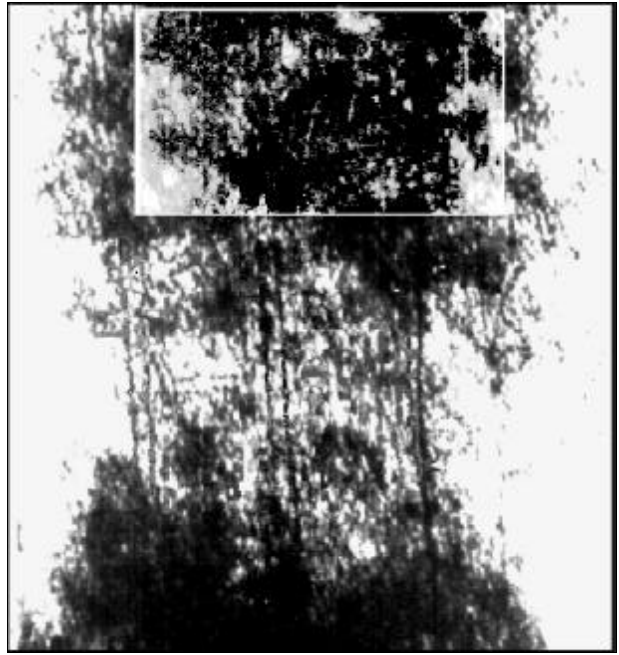

Figure 10. The «stripped» neck of a Z-pinch (major conditions are similar to those of Fig. 8; Z-pinch axis is directed vertically). Negative, t = 0 ns, time exposure 2 ns; image's height 1.65 cm. Diameter d of vertical tubules is ~0.3 mm, while for thinner tubules of various directions, including horizontal ones, d ~ 0.1-0.2 mm. Diameter of coaxial tubules seen, e.g., in the right hand side, is ~1 mm. The picture illustrates the presence of a network built up by the tubular rigid-body filaments which may be hidden in the ambient plasma at implosion and stagnation stages of the discharge but appear to be stripped by the magnetic field when it pushes the plasma out of the Z-pinch's neck (it is such event that leads to a singularity of the total electric current through the Z-pinch).

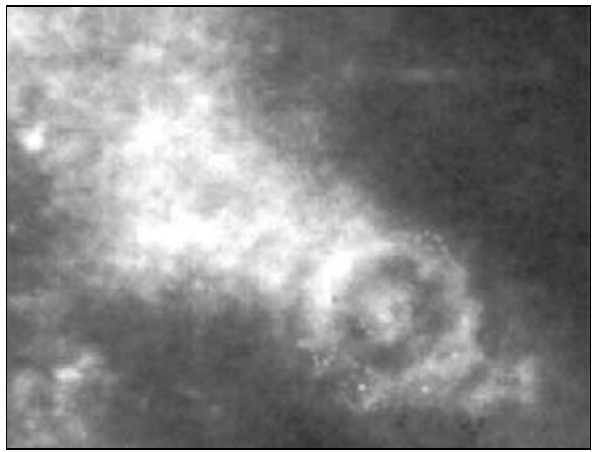

Figure 11. Typical tubular formation (of diameter ~ 0.5 mm) directed nearly perpendicular to the curvilinear filamentary current sheath at the implosion phase of the discharge in the plasma focus facility [7] of the Filippov type (for experimental conditions see also [8]). The image (positive) is a shadowgram taken from the facility's top (laser pulse of 2 ns duration passes up through a hole in the mushroom anode, i.e. the picture shows radial projection of the tubular formation). Image's width 2.5 mm.

### 3. MAJOR HYPOTHESES

The observed longevity of the pretty straight structures has lead to the conclusion [1,2] that the long-lived filaments should possess a microsolid skeleton. This assumes that only the quantum long-range bonds (like those in a condensed matter, rather than the interactions in a plasma composed of classical charged particles) may be responsible for the observed rigidity and survivability of LLFs. Besides, the connectivity of blocks (and the solidification of the structure on certain length scales) should be formed during electric breakdown stage and, hence, appear *prior* to formation of the (main) plasma itself.

The most probable candidate for the major microscopic building blocks of the LLF's skeletons was indicated, the *carbon nanotube* (which has for the first time been observed in 1991 by S. Iijima [9]) or similar nanostructures of other chemical elements. This choice [1,2] was determined by a number of physical-chemical properties of carbon nanotubes which, in combination, allow a new type of electric breakdown (not necessarily in a gas!) which is based on the build-up of skeletal structures from the microdust, either already present in the discharge chamber (due to its deliberate/undeliberate accumulation during, e.g., the «training» of the discharge chamber) or wildly formed during electric breakdown itself.

In order to resolve major difficulty in the above picture, namely the survivability of a condensed matter (i.e. skeletons) in a *hot* ambient plasma, of temperatures up to kiloelectronvolt range (in the latter point the situation substantially differs from physics of «dusty plasmas»), the microsolid skeletons were suggested [3(a,c,e)] to be self-protected from an ambient high-temperature plasma by a thin vacuum channels sustained self-consistently around the skeletons by the pressure of high-frequency (HF) electromagnetic waves, thanks to the skeleton-induced conversion of a small part of the incoming «static» magnetic field (poloidal, in tokamaks, or azimuthal, in Z-pinches) into HF waves of the TEM type (a «wild cable» model [3(a,c)]). Thus, the wild cable model has proposed common, and mutually interrelated, qualitative solution to the two problems: namely,

(i) survivability of skeletons, and

(ii) phenomenon of nonlocal (non-diffusive, in particular, ballistic) transport of energy (the latter was observed in last decade in various tokamaks, see e.g. the survey [10]).

The wild cable model was shown [3(a,c,d)] to be compatible with the measurements of HF electric fields in tokamak T-10, both inside [11] and outside [12] plasma column, and with similar measurements in a gaseous Z-pinch [13]. According to wild cable model, the TEM wave which propagates along the skeleton's straight section in a vacuum channel around this section may be a source of the intense plasma waves in the ambient plasma (therefore, one can extrapolate the scaling law of the TEM wave amplitude's spatial distribution from the channels around single straight section to the ambient plasma). It appears that the measured values [11,13] of the HF electric field averaged over space region with many wild cables, correspond to such amplitudes of the TEM wave which are sufficient for sustaining the vacuum channels of the width coinciding with that of the experimentally visible long straight sections of skeletons (for the case of a Z-pinch, see Figure 5 in [3(d)] for the radial profile of plasma density around straight section of a skeleton, which is calculated from a Poisson equation in the frame of a quasi-hydrodynamics of a plasma in a HF electric field).

## 4.    VERIFICATION OF HYPOTHESES

For verification of the above hypotheses the following proof-of-concept studies have been carried out which confirm the validity of the concept in several key points. The up-to-date results of these proof-of-concept studies are as follows.

Evidences for tubular structures in the range from several nanometers to several micrometers in diameter are found in the electronic (transmission and scanning) micrographs of various types of dust deposit (submicron and micron particles, and films, mostly carbon ones) in tokamak T-10 [14] (see Figs. 12-14 taken from [14]). A «wild» formation of nanotubular structures (i.e. their presence in the much wider range of experimental conditions than it was thought before) has been predicted in [1,2].

A bridge between the microdust of nanoscale size and the long-lived filaments in plasmas is drawn on the example of the structure which is essentially of non-hudrodynamic nature, namely the cartwheel-like structure (i.e. a ring with straight radial bonds on the axle). The topological similarity of the cartwheels in the above dust (namely, cartwheel of ~100 nm total diameter, assembled from tubules of ~10 nm diameter) and in the few centimeters size structures seen in the plasma images in small tokamaks [3(a,b)], is illustrated with Figures 14 and 15 taken from [14].

The trend of assembling bigger tubules from smaller ones may be seen in Figures 13 and 14 taken from [14]. This is compatible with hypothesis [1,2] which suggested tubular structures to be responsible for the self-similarity of LLFs. Besides, again in agreement with [1,2], it follows from available data that for all types of microdust skeletons of various topology and spatial dimensionality (namely fibers, spheres, balls, dendritic structures, etc.), contrary to quasi-continuous amorphous media, a tubular structure appears to be a key building block.

The presence of skeletons (tubules and cartwheels of millimeter-centimeter size) at initial stage of electric discharge was shown in various types of discharge: namely ,

• at «dark» stage of discharge in a vacuum spark [15] (electric current is less than 20% of its maximum, the plasma's self-emission is not yet detectable by the high-sensitivity detectors, skeletons are observed with the help of a laser shadowgraphy, see Fig. 18);

• during electric breakdown (i.e., before electric current appearance) in the database from tokamak T-6 [16] (visible light imaging with the help of an electronic optical converter (EOC), see Fig. 16);

• during electric breakdown in the database from plasma focus (imaging with the help of an EOC), see Fig. 19.

The analysis [16] of capabilities of the visible light imaging of tokamak plasmas disclosed an important limitation imposed by plasma rotation. The frame imaging, with 15 μs time exposure, by an EOC at t ~300 μs *before* appearance of the plasma current revealed the presence of tubules and cartwheels of diameters about few-several centimeters. Similar imaging after appearance of the plasma current didn't allow the resolution of structures: 15 μs exposure appeared to be insufficient for a plasma with rotation velocities certainly higher than those at electric breakdown stage. However, imaging at quasi-stationary stage of discharge in another regime of EOC's operation, namely a strick camera regime, showed similar structures as seen at the «dark» stage in the framing regime (see Fig.17). It follows that the temporal resolution as high as ~ 1 μs is needed for a rotating tokamak plasma.

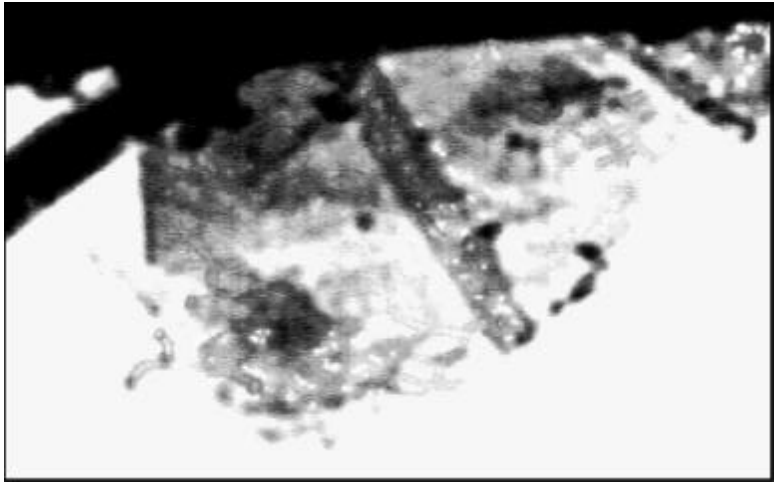

Figure 12. The transmission electron microscope (TEM) image (magnification 9,000) of a dust deposit at glass filter fibers. The dust was pumped out from a crimp in the tokamak T-10 chamber. Image's height 560 nm. Black region on the top of the figure is the image of a glass filter fiber. A large rod with the signs of the non-uniformity, which is probably caused by the inhomogeneity of its tubular sheath, is 370 nm long and of 55 nm diameter. This rod is connected with a network of substantially thinner tubules. Diameter of the tubule which enters large rod from the left, not far from the lower edge of the rod, is ~7 nm. The separate tubules are seen to the left from the above network.

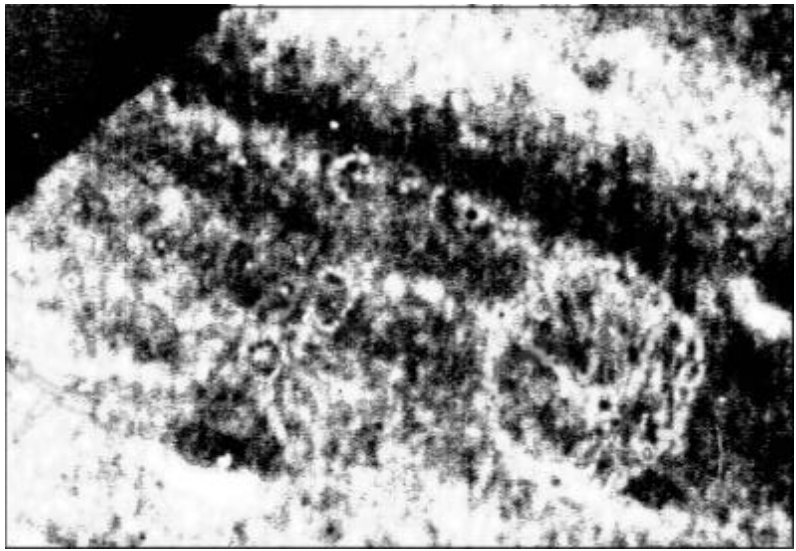

Figure 13. The scanning electron microscope (SEM) image (magnification 2,000) of a tubular formation in the surface layer of the film deposited at the internal surface of the T-10 tokamak chamber. Image's height 15 μm. Diameters of the tubule and the central spot on the tubule's edge are ~5 μm and ~1 μm, respectively.

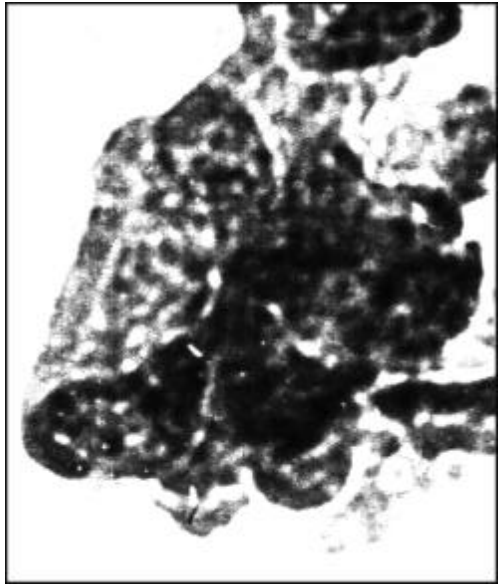

Figure 14. The transmission electronic microscope (TEM) image (magnification 34,000) of a part of the dust particle, of ~1.2 micrometer diameter, extracted from the oil used in the vacuum pumping system of tokamak T-10. Image's height 270 nm. The tubule whose edge with the distinct central rod is seen in the lower left part of the figure, is of ~70 nm diameter and ~140 nm long. Diameter of the slightly inhomogeneous cylinder which is seen on the left side of the tubule and is a constituent part of the tubule, is ~10 nm. The radial bonds between side-on cylinder and central rod are of ~10 nm diameter.

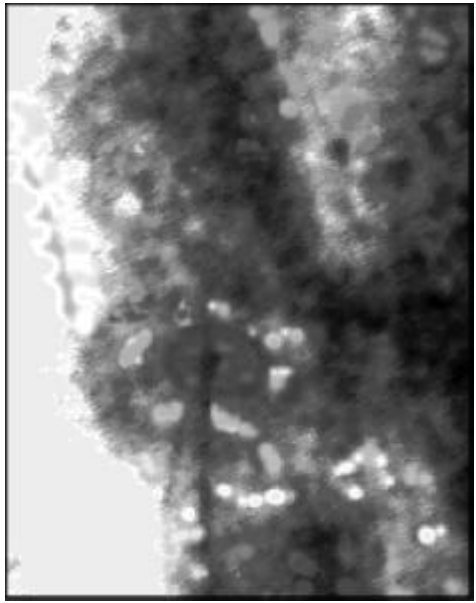

Figure 15. A cartwheel structure in the visible light image (positive) of plasma column near the axis (directed horizontally) of tokamak TM-2 (minor radius a = 8 cm). Image's height 5 cm. Diameters of larger and smaller ring-shaped structures on a common axle are ~ 2.2 cm and ~ 1 cm, respectively. Diameter of the axle at the cartwheel's plane is ~ 2 mm. Image is taken in visible light by a strick camera with effective time resolution < 1 μs (original image is taken from the database [6]).

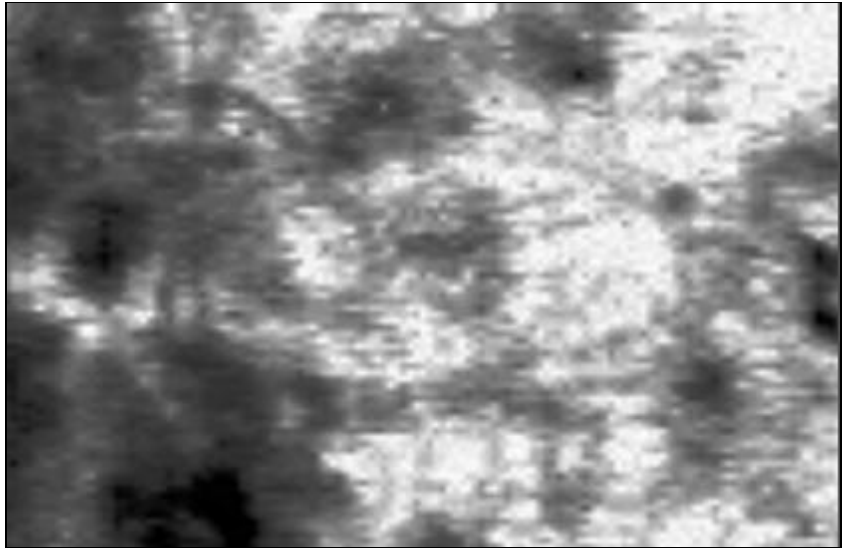

Figure 16. The elliptic image of a cartwheel-like structure (image's width ~ 2.5 cm, toroidal direction - horizontal) seen in tokamak T-6 at t ~300 μs *before* appearance of the plasma electric current. The image is taken by an electronic optical converter (EOC) in the framing regime (time exposure 15 μs).

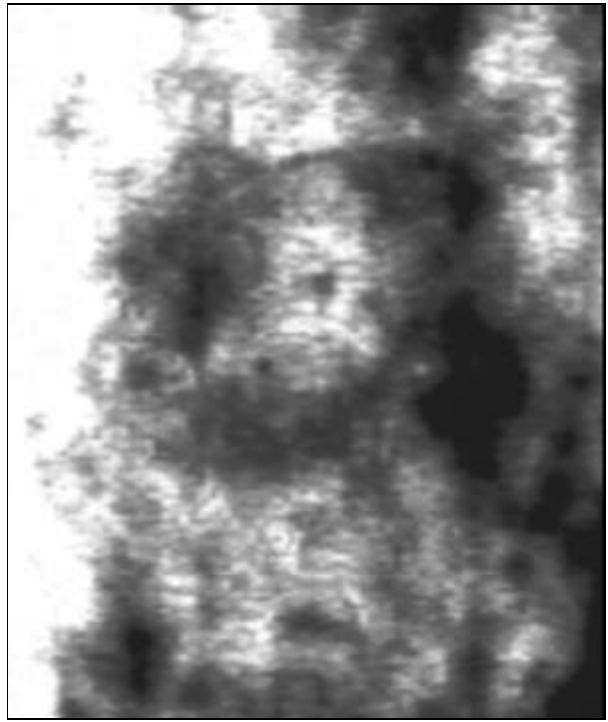

Figure 17. The elliptic image of a ring-shaped structure with a black central spot which is seen at the quasi-stationary stage of discharge in tokamak T-6. The image is taken in strick camera regime of EOC's operation, with effective time exposure ~ 1 μs. Image's width 2.5 cm. Large axis of the ellipse is 2.2. cm. There is also a tubular structure superimposed on the left part of the above ellipse (the lower edge of this tubule is of 6 mm diameter, and thick black spot in the center of this edge is of 2 mm size).

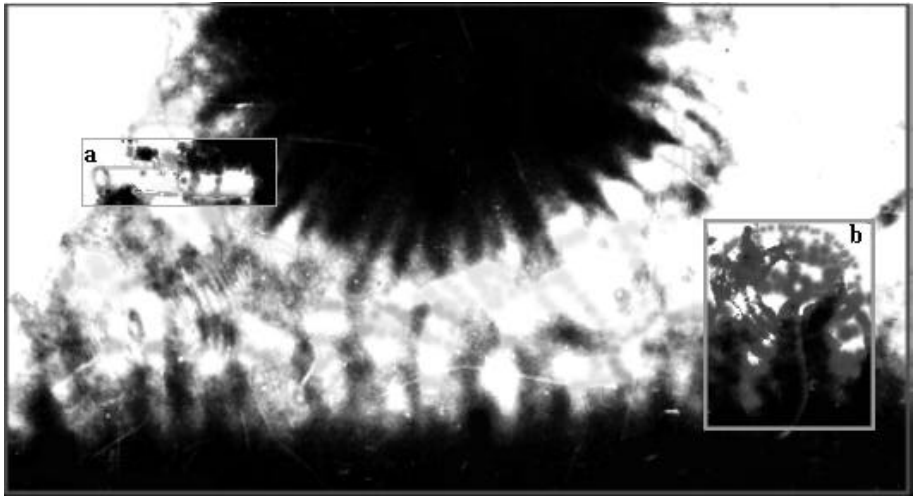

Figure 18. The tubular and cartwheel structures in a low-inductance vacuum spark (condenser capacitance 12 μF, voltage bias 10 kV, maximum current ~ 150 kA, period ~ 5 μs, flat cathode with central hole of diameter 3 mm is 2 mm from a round-shaped edge of a rod anode). The image is taken by a laser shadowgraphy (pulse duration 6 ns, λ=337 nm) using an electronic optical converter (EOC), at initial, «dark» stage when plasma's self emission is not yet detectable by the EOC (at this stage the electric current is lower than 20% of its maximum). The images in the windows 'a' and 'b' are processed with a higher level of contrasting. The cartwheels are seen in the window 'b' and, as an elliptic structure of larger size, in the left hand side of the image.

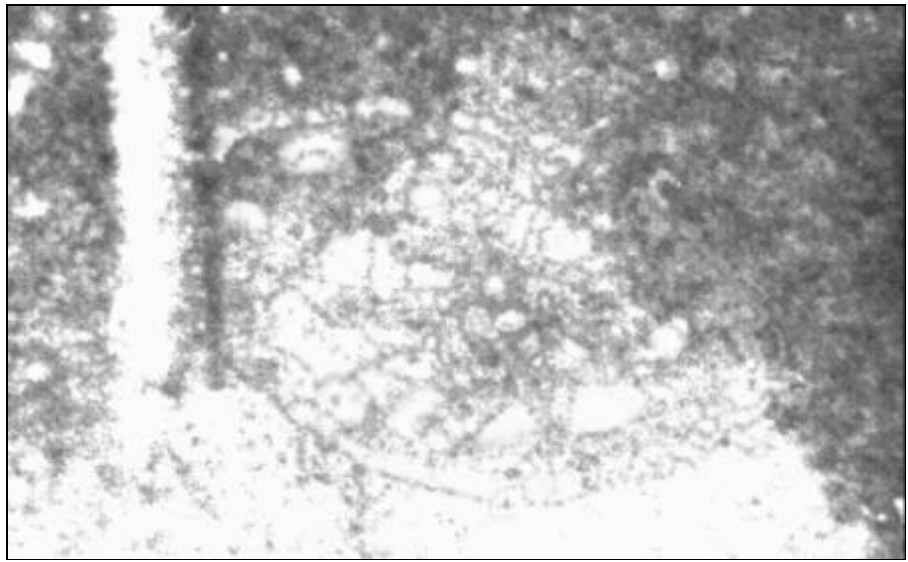

Figure 19. The tubular and cartwheel structures which are observed at time 100 ns before appearance of the electric current in the plasma focus facility [7] of the Filippov type (imaging with the help of an EOC; positive; time exposure 2 ns; image's width 4.6 cm). The structures are seen on the background of the annular vertical porcelain insulator whose left edge is seen at the left hand side of the image as a vertical white band (the mushroom-shaped anode is located upper than image's top, and the bottom of the cathode chamber is near the image's bottom).

## 5.   SUMMARY

The results of proof-of-concept studies very briefly reported in Sec. 4 give support, first of all, to the major point of the hypothesis [1,2], namely the necessity to go beyond the frame of _classical_ electrodynamics in describing the _long-range_ bonds in laboratory and cosmic plasmas. Here, we have to refer, first of all, to the phenomenon of the cartwheel-like structures that can not be explained in the frame of conventional approaches because of

(i) the presence of cartwheels in definitely «quantum» media (namely, in condensed matter – specifically, in various dust deposits), and especially, existence of cartwheels as separate solid blocks (see Fig. 14 and more examples in [14]);

(ii) observed conservation of essentially «non-classical» topology of cartwheels (namely, its incompatibility with the known hydrodynamic or particle beam structuring) in a broad range of plasma conditions, like e.g. geometry of the facility, for a wide range of experimental facilities;

(iii) rich statistics of observations that involve, e.g., practically every well-done plasma experiment with a proper high-resolution diagnostics and subsequent processing of plasma images, etc.;

(iv) insensitivity to specific way of imaging.

Further, the verification analyses (Sec.4) give support to the particular way towards quantum long-range structures suggested in [1,2], namely to the key role of _nanotubular_ blocks (specifically carbon nanotubes or similar nanostructures of other chemical elements) in composing the fractal solids which take the form of _self-similar skeletons_. The above self-similarity allows to draw a bridge from the nanotubular structuring in dust deposits to the macroscopic, of centimeter range size, long-lived structures in fusion plasmas.

It is important to note that the picture of potential implications of the hypothesis [1,2] outlined in the Summary of the survey [1] holds its form both for laboratory and space. In particular, the phenomenology of observations of skeletal structures does not contradict the phenomenology of the existing successful experiments (e.g., skeletons may be responsible for the observed stability of many plasmas and respective suppression of numerous instabilities predicted by the theory). Moreover, one can find new opportunities for resolving existing difficulties in interpreting the current picture of fusion plasmas (e.g., in describing the observed phenomena of nonlocal transport in plasmas and, probably, achieving a non-empirical description of the energy lifetime in the present fusion plasmas).

**Acknowledgments**

The authors are indebted to their colleagues who kindly presented the original databases from former experiments: namely, V.M. Leonov, S.V. Mirnov & I.B. Semenov, K.A. Razumova, and V.Yu. Sergeev (from tokamaks T-6, T-4, TM-2, and T-10, respectively) for Figures 1-7; A.R. Terentiev (from the plasma focus facility [7]) for Figures 11 and 19. We highly appreciate valuable collaboration with B.N. Kolbasov & P.V. Romanov, A.S. Savjolov, and V.A. Krupin, in the papers [14,15,16], respectively.

Our special thanks to V.I. Kogan for his invariable support and encouragement.